\documentclass[amsmath,amssymb,aps,prl,twocolumn]{revtex4-1}
\usepackage{graphicx}
\usepackage{bm}
\usepackage{array}
\usepackage{amsmath}

\begin{document}


\title{Anomalous change in the de Haas-van Alphen oscillations of CeCoIn$_5$ at ultra-low temperatures}

\author{Hiroaki Shishido$^{1,2}$}
\email[]{shishido@pe.osakafu-u.ac.jp}
\author{Shogo Yamada$^3$}
\author{Kaori Sugii$^3$}
\author{Masaaki Shimozawa$^3$}
\author{Youichi Yanase$^4$}
\author{Minoru Yamashita$^3$}
\email[]{my@issp.u-tokyo.ac.jp}

\affiliation{
$^1$Department of Physics and Electronics, Graduate School of Engineering, Osaka Prefecture University, Sakai, Osaka 599-8531, Japan.\\
$^2$Institute for Nanofabrication Research, Osaka Prefecture University, Sakai, Osaka 599-8531, Japan.\\
$^3$The Institute for Solid State Physics, The University of Tokyo, Kashiwa, 277-8581, Japan\\
$^4$Department of Physics, Kyoto University, Kyoto 606-8502, Japan.
}

\date{\today}

\begin{abstract}
We have performed de Haas-van Alphen (dHvA) measurements of the heavy-fermion superconductor CeCoIn$_5$ down to 2 mK above the upper critical field.  
We find that the dHvA amplitudes show an anomalous suppression, concomitantly with a shift of the dHvA frequency, below the transition temperature $T_{\rm n}=20$ mK.
We suggest that the change is owing to magnetic breakdown caused by a field-induced antiferromagnetic (AFM) state emerging below $T_{\rm n}$,  revealing the origin of the field-induced quantum critical point (QCP) in CeCoIn$_5$.
The field dependence of $T_{\rm n}$ is found to be very weak for 7--10 T, implying that an enhancement of AFM order by suppressing the critical spin fluctuations near the AFM QCP competes with the field suppression effect on the AFM phase.
We suggest that the appearance of a field-induced AFM phase is a generic feature of unconventional superconductors, which emerge near an AFM QCP, including CeCoIn$_5$, CeRhIn$_5$, and high-$T_{\rm c}$ cuprates.
\end{abstract}

\pacs{}
\maketitle

The competition between distinct quantum states at a quantum critical point (QCP), where the second-order transition temperature is suppressed to absolute zero, prevents the ground state from attaining either state and enhances quantum fluctuation in the vicinity of the QCP \cite{Sachdev2010}. The fate of electronic states influenced by the enhanced quantum fluctuations near a QCP has been a central issue of modern physics. Many anomalous states such as non-Fermi liquid (NFL) behavior and unconventional superconductivity have been observed near QCPs in various materials including high-$T_{\rm c}$ cuprates \cite{Sachdev2010}, iron-pnictides \cite{Shibauchi2014}, and heavy fermions (HFs) \cite{Gegenwart2008}. These unconventional superconductors often show a maximum transition temperature near the QCP, suggesting that the enhanced quantum fluctuation is the key to understanding their superconductivity.

HF systems have emerged as prototypical systems for studying QCPs, because the strong mass renormalization that occurs through hybridization of $f$-electrons with conduction electrons lowers the relevant energy scale on which the effects take place. Thus, the ground state can be easily tuned at experimentally accessible pressures or magnetic fields. In particular, unconventional superconductivity in HFs has been most extensively studied in CeCoIn$_5$ because a $d$-wave superconducting state emerges at ambient pressure without chemical substitution \cite{Izawa2001}. The $d$-wave superconductivity has been shown to be located in the vicinity of an antiferromagnetic (AFM) QCP \cite{Kawasaki2003,Tokiwa2013}. Various measurements have further revealed a crossover of NFL behavior at zero field to FL behavior at high fields, indicating the presence of a field-induced QCP near the upper critical field, $H_{\rm c2}$ \cite{Bianchi2003,Paglione2003,Howald2011,Zaum2011}.

Despite the accumulating evidence for field-induced QCP, no AFM state corresponding to the QCP has been observed outside the superconducting phase. This apparent absence has been attributed to the AFM state being hidden at an inaccessible negative pressure \cite{SarraoThompson2007} or superseded by the superconductivity \cite{Tokiwa2013}. For $H // ab$, a spin-density-wave order is induced inside the superconducting phase near $H_{\rm c2}$, which is recently discussed in terms of a condensation of the spin resonance in the superconducting phase \cite{Stock2012, Raymond2015}.
 This coexisting ``Q-phase'' \cite{Kenzelmann2008}, however, vanishes when the field is tilted from the $ab$ plane \cite{Correa2007} whereas the NFL behaviors are observed regardless of the field direction. Thus, it is necessary to search for an ordered state for $H // c$ at lower temperatures to clarify the origin of the field-induced QCP and its interplay with the unconventional superconductivity of the material.

In this Letter, we report de Haas-van Alphen (dHvA) measurements for $H // c$ down to 2 mK using our home-made nuclear-demagnetization cryostat. 
We find that the dHvA amplitudes deviate from the conventional Lifshitz-Kosevich (LK) formula~\cite{Shoenberg} and show an anomalous decrease with a shift of the dHvA frequency below a transition temperature $T_{\rm n}$, suggesting an emergence of the putative AFM state.

High-quality single crystals of CeCoIn$_5$ were grown by the In-flux method~\cite{Shishido2002}. Measurements of the magnetic torque were performed using a capacitance cantilever technique up to 10 T.
The lowest temperature of the cryostat was measured by a melting curve thermometer calibrated by the transition points of $^3$He~\cite{Greywall1985}. To ensure the lowest temperature of the samples, the samples were immersed in liquid $^3$He of which the temperature was monitored by a vibrating wire thermometer~\cite{Carless1983} placed in the same liquid $^3$He (see Supplemental Material (SM)~\cite{SM} for details).

The dHvA oscillation for $H // c$ and the corresponding fast Fourier transformation (FFT) spectrum are shown in Figs. 1(a) and 1(b), respectively. The fundamental branches are assigned as $\alpha_1$, $\alpha_2$, $\alpha_3$, $\beta_2$, $\varepsilon$, and $\gamma$ as indicated in Fig. 1(b), which agree with those of the previous report \cite{Settai2001}. 
As shown in Fig. 1(b), the dHvA amplitudes of the all branches increased as temperature was lowered to 20 mK in accordance with the LK formula \cite{Shoenberg}. 
The temperature dependence of the dHvA amplitudes of all $\alpha$ branches above 20 mK can be well fitted by the LK formula (the solid lines in Fig. 1(c)), enabling us to estimate the effective cyclotron mass $m_{\rm c}^*$ for each $\alpha$ branch which is also in good agreement with the previous report~\cite{Settai2001}. 
However, the dHvA amplitudes of all $\alpha$ branches deviated from the LK formula  below $\sim 20$~mK and decreased as lowering the temperature (Fig. 1(c)).
This anomalous decrease was also observed in $\gamma$ branch but not in $\varepsilon$ branch \cite{SM}.

\begin{figure}[tbh]
\includegraphics[width=0.8\linewidth]{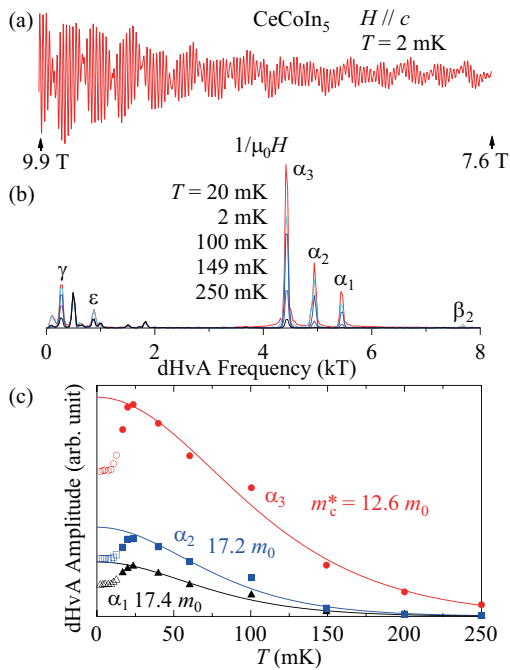}	
\caption{
Quantum oscillations of CeCoIn$_5$. 
(a) dHvA oscillation at 2 mK for 7.6--9.9 T after subtracting the background signal. 
(b) FFT spectra of the dHvA oscillations obtained in the same field range as (a) at 20, 2, 100, 149, and 250 mK in descending order of the signal size.
Note that the spectrum at 2 mK is smaller than that at 20 mK. See the main text and SM for details \cite{SM}.
(c) The temperature dependence of the dHvA amplitudes of $\alpha_1$(black triangles), $\alpha_2$ (blue squares), and $\alpha_3$ (red circles) obtained in the same field range of (a). 
The open (filled) data were taken by using nuclear demagnetization (dilution refrigeration).
The solid lines show fits for the data above $T_{\rm n}$ from the standard Lifshitz-Kosevich formula \cite{Shoenberg}. 
The cyclotron effective mass ($m_{\rm c}^*$) estimated from the LK fit for each $\alpha$ branch is indicated with the data. 
}
\label{fig1}
\end{figure}

To analyze the field dependence in detail, the dHvA signal of $\alpha_3$ branch, which showed the largest signal, was isolated by a steep bandpass filter.
The temperature dependence of the dHvA amplitude at different field strengths is shown in Fig.~2(a).
The decrease of the amplitude below $T_{\rm n}$ was most clearly observed at $\sim 8$ T and became less pronounced at higher field strengths. 
Below 7.4 T, $T_{\rm n}$ slightly decreased  and the decrease in the dHvA amplitudes became smaller. The decrease was not clearly resolved at 6.0 T because of the small dHvA signal.

Further, we analyzed the temperature dependence of the dHvA frequency by using a phase shift analysis \cite{Shoenberg}.
The phase shift of a dHvA oscillation, $\Delta P(T,H) = P(T,H) - P(T_0, H)$, is proportional to the shift of the dHvA frequency $\Delta F(T,H) = F(T,H) - F(T_0, H)$ as, $\Delta P(T,H)/P(T_0, H) = \Delta F(T,H)/F(T_0, H)$, where $P(T,H)$ is the peak field of the dHvA oscillation and $T_0$ is a reference temperature.
As shown in Fig.~2(b), the temperature dependence of $\Delta F(T,H)$ is found to show a kink at $T\sim T_{\rm n}$, which was followed by a slight decrease of the frequency at lower temperatures.

A similar suppression of the dHvA amplitude of $\alpha_3$ has been reported below ~100 mK and at 13--15 T \cite{McCollam2005}, which was discussed in terms of a strong spin dependence of the effective mass. 
We applied the spin-dependent LK formula to our results at various fields (the blue dashed lines in Fig.~2(a)). 
As shown in Fig.~2(a), the spin-dependent LK formula reasonably reproduces the data at 9.5 T, which is consistent with the previous report at higher fields~\cite{McCollam2005}.
However, the spin-dependent LK formula clearly fails to reproduce the rapid decrease of the dHvA signal below $T_{\rm n}$ observed for 7.6--9.0 T, even by assuming a very large difference between the effective mass of the spin-up electrons and that of the spin-down ones (see SM~\cite{SM} for details of these fittings).
Moreover, this model cannot explain the change in the dHvA frequency below $T_{\rm n}$.
These results indicate that a drastic change in the electronic state of the material occurs below $T_{\rm n}$, leading us to suggest that the anomalous change in the dHvA amplitudes corresponds to a phase transition. 

\begin{figure}[htb]
\includegraphics[width=0.8\linewidth]{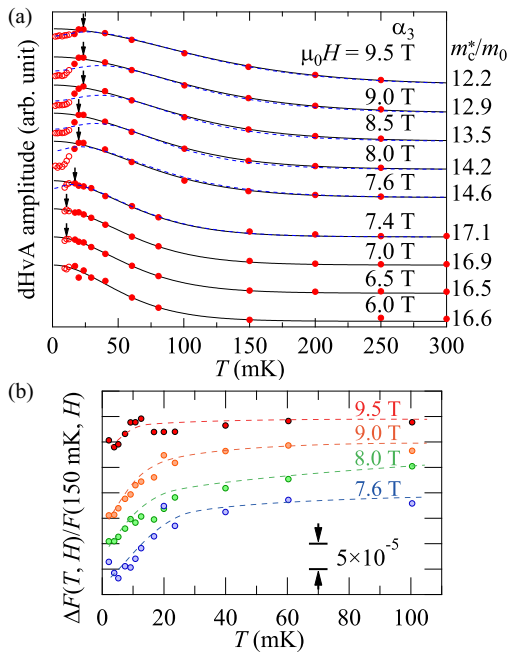}	
\caption{
(Color online)
(a) Temperature dependence of the dHvA amplitudes of $\alpha_3$ at different fields. The data are shifted for clarity. 
The open (filled) data were taken by using nuclear demagnetization (dilution refrigeration).
A clear suppression of the dHvA amplitude from the standard Lifshitz-Kosevich (LK) formula (solid lines, $m^*_c/m_0$ used for the fits are shown right) \cite{Shoenberg} was observed.
The transition temperature $T_{\rm n}$ is determined as the onset temperature where the dHvA amplitude starts to deviate from the LK formula (shown by the arrows).
The blue dashed lines show the fits for the spin-dependent LK formula \cite{McCollam2005} (see the main text).
The saturation observed at the lowest temperature may be caused by a saturation of the AFM energy gap or non-equilibrium of the sample temperature (see SM~\cite{SM} for details).
(b)
The temperature dependence of the normalized shift of the dHvA frequency $\Delta F(T,H) /F (\textrm{150 mK}, H)$ of $\alpha_3$. The data are shifted for clarity. The dashed lines are guides to the eye.
}
\label{fig2}
\end{figure}

The revised $H$--$T$ phase diagram of CeCoIn$_5$ with the field-induced phase is shown in Fig. 3(a). 
We found that the field dependence of $T_{\rm n}$ was very weak for 7--10 T, although we have to note that determining $T_{\rm n}$ has a large ambiguity of $\sim20$\% because the dHvA measurements were performed by sweeping fields at a constant temperature.
We have observed similar anomalies of the dHvA amplitude in a different single crystal \cite{SM}.
An anomalous reduction has also been observed in the magneto resistance at $\sim 20$ mK and at 8 T \cite{SM, Howald2011}.
It should also be noted that the absence of the anomaly at low field is consistent with the previous work done at 6--7 T \cite{McCollam2008, SM}.

Here, we suggest that a field-induced AFM order provides the most plausible explanation for the anomaly observed in our dHvA measurements. 
An AFM transition in the simple tetragonal structure of CeCoIn$_5$ ($P$4/$mmm$) modifies the dHvA frequencies by a folding of the Brillouin zone.
However, the paramagnetic Fermi surface can still be observed in the AFM phase with a larger attenuation by magnetic breakdown \cite{Shoenberg}, as observed in NdIn$_3$~\cite{SETTAI1994}.
The magnetic breakdown probability ($P_{MB}$) is given by $P_{MB}=\exp (-\epsilon_g^2 / \hbar \omega_c \epsilon_F)$, where $\epsilon_g$ is the energy gap for the magnetic breakdown, $\omega_c$ is the cyclotron frequency, and $\epsilon_F$ is the Fermi energy \cite{Shoenberg}. Because $T_{\rm n}$ is two or three orders magnitudes smaller than that of typical 4$f$-electron AFM materials (where the AFM transition typically takes place at a few K or higher e.g. $\sim 6$ K for NdIn$_3$~\cite{SETTAI1994}), the magnetic moment would be much smaller than these AFM materials, resulting in a tiny energy gap at the magnetic zone boundary. As a result, majority of electrons undergo magnetic breakdown and the new dHvA branches from the folded Brillouin zone in the AFM phase may not be observable within our experimental accuracy.
This explains the suppression of the dHvA amplitudes below $T_{\rm n}$ without new dHvA signals from the folded Brillouin zone.

The appearance of the AFM order is also supported by the slight change of the dHvA frequency below $T_{\rm n}$ (Fig. 2(b)).
It is known that the measured dHvA frequency $f_m$ is given by the zero-temperature extrapolation of the true dHvA frequency $f_t$ as $f_m = f_t - H \left( \partial f_t / \partial H \right)$.
Either the change of the slope $\left( \partial f_t / \partial H \right)$ and/or the Fermi surface size $(\propto f_t) $ cause a change of $f_m$.
Even if magnetic breakdown occurs, $f_m$ can be modified by a change of the slope of $f_t$ caused by the AFM ordering.
Therefore, althought the change in the dHvA frequency is very small, the field dependence of the dHvA frequency is consistent with the field-induced AFM order.

The absence of the anomalous suppression in $\varepsilon$ branch \cite{SM} is also consistent with the AFM transition because $\varepsilon$ branch corresponds to a  small pocket located at the zone center \cite{Settai2011} and thus is hardly affected by the band folding.
In addition, a kink in the temperature dependence of the resistivity at 8 T has been observed at temperatures very close to $T_{\rm n}$ (see the left column of Fig. 3 in ref. \cite{Howald2011}). Although the origin of the kink is not discussed in ref. \cite{Howald2011}, the kink can be consistent with a reduction of the magnetoresistance in the AFM phase. 
Therefore, we conclude that the anomalous change of the dHvA amplitudes below $T_{\rm n}$ is most consistent with the emergence of an AFM phase. We note that splitting of the dHvA frequency typically expected for an AFM phase was not observed simply because our field range of 6--10 T was too narrow to allow the detection. 
A change of the torque signal by AFM order was also not resolved at $T_{\rm n}$, which was probably hindered by the change of the dHvA signal.
The possibilities of multipole ordering, Lifshitz transitions and nuclear spin ordering can be safely excluded as described in SM~\cite{SM}.

\begin{figure*}[!thb]
\includegraphics[width=0.9\linewidth]{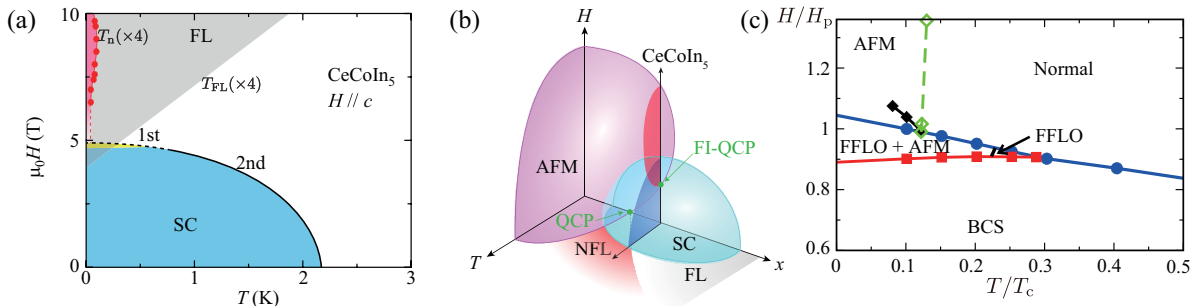}
\caption{
(Color online) Phase diagrams of CeCoIn$_5$. 
(a) $H$--$T$ phase diagram. Field-induced phase found by our measurements (pink), the Fermi liquid (FL) region (gray, taken from ref. \cite{Zaum2011}), the superconducting (SC) phase (blue), and the high-field SC phase (yellow) \cite{Kumagai2006} are shown. Both $T_{\rm n}$ (red circles) and $T_{\rm FL}$ are multiplied by 4 for clarity. The black dashed and solid lines represent the first- and the second-order SC transitions, respectively.
(b) A schematic $H$--$T$--$x$ phase diagram near the AFM QCP, where $x$ denotes pressure or chemical substitution. 
The cross section corresponds to the $H$--$T$ phase diagram of CeCoIn$_5$, in which a presumed field-induced QCP (FI-QCP) is also shown at the intersection with the AFM boundary.
(c) A calculated phase diagram near $H_{\rm c2}$ by a mean field approximation (solid symbols and solid lines) \cite{Yanase2009} and a FLEX approximation (open symbols and dashed line) \cite{Yanase2008}. 
The temperature and the magnetic field are normalized by the transition temperature $T_{\rm c}$ and the Pauli limiting field $H_{\rm p} = 1.25 T_{\rm c}$, respectively. 
The lines are guides to the eye (see SM~\cite{SM} for details).
}
\label{fig3}
\end{figure*}

The revised $H$--$T$ phase diagram of CeCoIn$_5$ (Fig. 3(a)) reveals that the field-induced AFM phase is located at the boundary of the unconventional superconductivity. A similar $H$--$T$ phase diagram has been observed in the sister compound CeRhIn$_5$ where an AFM ground state at ambient pressure changes to a superconducting state under pressure \cite{SarraoThompson2007,Park2006}. The pressure dependence of the $H$--$T$ phase diagram of CeRhIn$_5$ can be summarized in a schematic $H$--$T$--$x$ phase diagram (Fig. 3(b)) where $x$ denotes pressure for CeRhIn$_5$. Given that the field-induced phase is observed at very low temperature, the $H$--$T$ phase diagram of CeCoIn$_5$ at ambient pressure may be considered to be a cross section at the vicinity of the AFM QCP in the $H$--$T$--$x$ phase diagram. Thus, CeCoIn$_5$ is a prominent superconductor where the interplay of unconventional superconductivity, magnetic order, and non-Fermi liquid behaviors near the QCP can be studied without ambiguity caused by the application of pressure or chemical doping. 
Such $H$--$T$--$x$ phase diagrams ($x$ = pressure or chemical substitution) have not only been observed in HFs \cite{Gegenwart2008,SarraoThompson2007,Park2006, Settai2011} but also in high-$T_{\rm c}$ cuprates \cite{Sachdev2010,Lake2002}. These similarities suggest that the $H$--$T$--$x$ phase diagram is generic to unconventional superconductors, which emerge in the vicinity of an AFM QCP, including CeCoIn$_5$.

Remarkably, the transition temperature of the field-induced AFM phase depended on field only weakly for 7--10 T even though the Zeeman energy at 10 T is about three orders of magnitude larger than $k_B T_{\rm n}$.
This weak field dependence implies that an enhancement effect on AFM order by the magnetic field competes with the suppression effect.
If the critical spin fluctuation near the QCP \cite{Kawasaki2003,Tokiwa2013,Bianchi2003,Paglione2003,Howald2011,Zaum2011} suppresses AFM order, the magnetic field suppresses the critical spin fluctuation, giving rise to an enhancement effect on AFM order.
To take into account the spin fluctuation effect near an AFM QCP, we examined the AFM transition line in the normal state of CeCoIn$_5$ by adopting the fluctuation exchange (FLEX) approximation \cite{Yanase2008}. We also calculated the $H$--$T$ phase diagram in the superconducting phase of CeCoIn$_5$ with a neighboring AFM phase by the mean field approximation \cite{Yanase2009} and plot them in Fig. 3 (c).
Whereas the mean field method can reproduce the first order superconducting transition near $H_{\rm c2}$, the critical spin fluctuation can be included only in the FLEX method.
As shown in Fig. 3 (c), the AFM transition temperature calculated by the FLEX approximation (dashed line) is found to increase as field increases, in contrast to the transition line with a negative slope by the mean field approximation (solid line above $H_{\rm c2}$). This difference demonstrates the field enhancement effect on the AFM transition temperature by suppressing the critical spin fluctuation as consistent with our results.
In fact, such a transition line has been reported in CeIrSi$_3$ near an AFM QCP under pressure of $\sim2.3$ GPa~\cite{Settai2011}, where the Zeeman energy is also much larger than the energy scale of the AFM transition, suggesting that the field independence of the transition line is a common feature near an AFM QCP.

We suggest that the field-induced AFM phase is possibly related to the high-field superconducting (HFSC) phase for $H // c$, which has been discussed as a spatially inhomogeneous superconducting state \cite{Kumagai2006} termed as a Fulde--Ferrell--Larkin--Ovchinnikov (FFLO) state \cite{FF64,LO65}. In the case of $d$-wave superconductivity located in the vicinity of an AFM QCP, such as that found in CeCoIn$_5$, calculations by the FLEX approximation have indicated that an FFLO state is stabilized \cite{Yanase2008}, which is also consistent with NMR measurements \cite{Kumagai2006}. For $H // ab$, AFM order coexists with the superconductivity in the Q-phase \cite{Kenzelmann2008}. This Q-phase has also been discussed in terms of an FFLO state because an FFLO state enhances AFM order through the appearance of an Andreev bound state localized around the gap nodes in real space  and by coupling between AFM order and pair-density-wave \cite{Yanase2009,Agterberg2009}. 
As shown in Fig.~3 (c), the transition temperature of the AFM phase increases in the FFLO phase. Therefore, we speculate that, if a field-induced AFM phase is also hidden at ultra-low temperatures for $H // ab$, the AFM phase is enhanced in the FFLO state, as shown in Fig. 3(c), and is observed as the Q-phase. 
Confirming AFM order for both $H // c$ and $H // ab$ and identifying the $q$ vector by local probe measurements such as nuclear magnetic resonance or muon spin resonance will be important future issues to clarify the relation between these phases.

In summary, we have observed anomalous changes in the dHvA oscillations in the normal state of CeCoIn$_5$ below $T_{\rm n}=20$ mK. We attribute these anomalies to the emergence of an AFM state which is the origin of the AFM QCP in CeCoIn$_5$. We suggest that CeCoIn$_5$ shares a phase diagram with other materials near an AFM QCP. We also suggest that AFM order is enhanced by suppressing the critical spin fluctuations in the vicinity of the AFM QCP in high field, which gives rise to the weak field dependence of $T_{\rm n}$ as supported by the FLEX calculation.
We believe that the development of the dHvA measurements under ultra-low temperature has extensive potential to shed a new light on unexplored phenomena at ultra-low temperatures.

\begin{acknowledgments}
We thank D. Aoki, S. Kambe, Y. Matsuda, Y. Matsumoto, Y. $\mathrm{\bar{O}}$nuki, H. Sakai, T. Shibauchi, Y. Tada, and H. Tokunaga for discussions. This work was performed under the Visiting Researcher's Program of the Institute for Solid State Physics, University of Tokyo, and was supported by the Toray Science Foundation, and KAKENHI (Grants-in-Aid for Scientific Research) Grant Numbers 15K05164, 15H05745, 15H05884, 15K17691, 16H00991, 16K05456, 16K17742, 16K17743, and 17K18747.
\end{acknowledgments}


\providecommand{\noopsort}[1]{}\providecommand{\singleletter}[1]{#1}%

\end{document}